**Maximizing Specific Loss Power for Magnetic Hyperthermia by Hard-Soft Mixed Ferrites**

Shuli He[1,2,4], Hongwang Zhang[2], Yihao Liu[1,3], Fan Sun[2], Xiang Yu[1], Xueyan Li[1], Li Zhang[1], Lichen Wang[1], Keya Mao[3], Gangshi Wang[3], Yinjuan Lin[3], Zhenchuan Han[3], Renat Sabirianov[5], and Hao Zeng[2*]

[1] Department of Physics, Capital Normal University, Beijing 10048, China

[2] Department of Physics, University at Buffalo, SUNY, Buffalo, New York 14260, USA

[3] Chinese PLA General Hospital, Beijing 10048, China

[4] Beijing Advanced Innovation Center for Imaging Technology, Beijing 10048, China

[5] Department of Physics, University of Nebraska-Omaha, Omaha, NE 68182, USA

**Abstract**

We report maximized specific loss power and intrinsic loss power approaching theoretical limits for AC magnetic field heating of nanoparticles. This is achieved by engineering the effective magnetic anisotropy barrier of nanoparticles via alloying of hard and soft ferrites. 22 nm $Co_{0.03}Mn_{0.28}Fe_{2.7}O_4/SiO_2$ NPs reached a specific loss power value of 3417 W/$g_{metal}$ at a field of 33 kA/m and 380 kHz. Biocompatible $Zn_{0.3}Fe_{2.7}O_4/SiO_2$ nanoparticles achieved specific loss power of 500 W/$g_{metal}$ and intrinsic loss power of 26.8 nHm$^2$/kg at field parameters of 7 kA/m and 380 kHz, below the clinical safety limit. Magnetic bone cement achieved heating adequate for bone tumor hyperthermia, incorporating ultralow dosage of just 1 wt% of nanoparticles. In cellular hyperthermia experiments, these nanoparticles demonstrated high cell death rate at low field parameters. $Zn_{0.3}Fe_{2.7}O_4/SiO_2$ nanoparticles show cell viabilities above 97% at concentrations up to 500 μg/ml within 48 hrs, suggesting toxicity lower than that of magnetite.

Magnetic nanoparticles (NPs) have received great attention over the past several decades due to their potential biomedical applications in targeted drug delivery, biological separation, magnetoresistive bio-sensing, magnetic resonance imaging, and as heat dissipation agents in gene transcription, neural stimulation and cancer treatment.[1-12] Magnetic hyperthermia, first proposed by Gilchrist in 1957,[13] employs heat dissipation by magnetic NPs in an alternating current (AC) magnetic field to kill tumor cells. The design and synthesis of magnetic NPs should take into consideration the following constraints: first of all, they should have the highest possible specific loss power (SLP) within the field and frequency range deemed safe for human body to avoid potential side effects and to be useful for treatment of small tumors;[11] second, they should be close to superparamagnetic (SPM) with low magnetostatic interactions to avoid agglomeration;[14,15] and third, they should be biocompatible with low cytotoxicity.[16-19]

Iron oxide NPs are the most commonly used materials in magnetic hyperthermia because of their low toxicity.[20-23] Other ferrite NPs such as manganese ferrite, and more recently, zinc ferrite have also been explored due to their high magnetization among the ferrite family and stability against oxidation.[24,25] Relatively high SLP values have been achieved by these NPs.[26] However, further increasing the SLP by increasing the saturation magnetization ($M_S$) would be futile. Although a high $M_S$ is beneficial for increasing SLP, high $M_S$ materials are typically metallic and face stability and toxicity issues in physiological environment.[21] An alternative approach to maximizing the SLP is to tune the effective anisotropy of the NPs. For example, shape anisotropy can be used to increase SLP of iron oxide nanocubes.[20] Very large SLP has been obtained by tuning the anisotropy of NPs through hard-soft exchange-coupled core/shell NP approach.[27] However, they were achieved at high field amplitude and frequency values unsuitable for clinical applications.[27]

In this work, we report the design and synthesis of monodisperse SPM NPs with maximized SLP and intrinsic loss power (ILP) at different field parameters. ILP is defined as SLP/$H^2f$ under linear response theory to compare the performance of NPs measured under different field parameters.[28] We first show that SLP can be maximized at $H$ = 33 kA/m and $f$ = 380 kHz, by alloying of hard cobalt ferrite and soft manganese ferrite to make $Co_xMn_{(0.3-x)}Fe_{2.7}O_4$, and tuning the size and composition of the mixed ferrite NPs. Unlike the core/shell approach, alloying allows convenient control of the effective anisotropy independent of NP size. A thin silica shell coating renders water-solubility and bio-functionality.[22] 22 nm $Co_{0.03}Mn_{0.28}Fe_{2.7}O_4/SiO_2$ NPs reach a SLP value of 3417 W/$g_{metal}$. We further optimize the composition of biocompatible $Zn_xFe_{3-x}O_4$ NPs for enhanced SLP under clinically safe field parameters (the product of field amplitude and frequency $Hf$ < 5×10$^9$ A/(m·s)). The $Zn_{0.3}Fe_{2.7}O_4$ NPs achieved SLP and ILP values of 1010 W/$g_{metal}$ and 15.7 nHm$^2$/kg at $H$ = 13 kA/m, 500 W/$g_{metal}$ and 26.8 nHm$^2$/kg at 7 kA/m, and 282 W/$g_{metal}$ and 59.9 nHm$^2$/kg at 3 kA/m, respectively ($f$ = 380 kHz). $Zn_{0.3}Fe_{2.7}O_4/SiO_2$ NPs show no cytotoxicity after 48 hrs at concentrations up to 500 μg/ml. Using the optimized biocompatible NPs, we achieved comparable temperature rise with significant decrease in dosage in a model mimicking bone tumor hyperthermia.[29,30] Cellular hyperthermia using 300 μg/ml $Zn_{0.3}Fe_{2.7}O_4/SiO_2$ NPs resulted in >89% cell death upon 10 min exposure to AC field of $H$ = 13 kA/m and $f$ = 380 kHz. Our designed biocompatible NP with maximized SLP and ILP provide a pathway towards targeted magnetic hyperthermia treatment of small tumors and metastases[11], and rapid remote neural stimulation.[4]

The mechanism of heat loss in magnetic hyperthermia is the energy dissipated during the magnetization reversal, and is therefore proportional to the product of the area of the AC magnetic hysteresis loop and the frequency.[31] An estimate of the upper limit of achievable SLP for ferrite NPs can be done as follows: $SLP = \alpha \frac{4\mu_0 M_S H_{max}}{\rho} f$, where $M_S$ is the saturation magnetization, $H_{max}$ the amplitude of the magnetic field, $\rho$ the mass density of the NPs and $f$ the frequency. $\alpha$ is a dimensionless factor describing the deviation from a square hysteresis, which is related to the degree of alignment, and should also dependent on the types of effective anisotropy (uniaxial vs cubic) and inter-particle interactions. Using $\alpha$ =1 for aligned Stoner-Wohlfarth particles with uniaxial anisotropy, $M_S$ ~ 480 kA/m, $H_{max}$ = 33 kA/m, $\rho$ ~ 5 g/cm$^3$ and $f$ =380 kHz, SLP is estimated to be ~ 5000 W/g$_{NP}$ or ~ 7000 W/g$_{metal}$. Any value higher than that is likely non-physical. For completely random ensembles, $\alpha$ is reduced to 0.39.[31] Note here the optimal AC coercivity ($H_c$) should be ~ $0.8H_{max}$. The AC hysteresis area is maximized this way despite the fact that a fraction of the NPs cannot be switched[31]. $\alpha$ as high as 0.46 has been reported for magnetosomes aligned in an external field.[32] Thus the limit of SLP for any ferrite NPs at a field of 33 kA/m and 380 kHz is about ½*7000 ~ 3500 W/g$_{metal}$.

How can we then tune the properties of NPs to approach the theoretical limits? The key is to engineer the anisotropy energy barrier for magnetization reversal to match a specific field amplitude and frequency for a particular application.[4,15-18] According to the Stoner-Wohlfarth model,[33] the anisotropy barrier is proportional to K$_u$V, where K$_u$ is the uniaxial anisotropy constant and V is the volume of the magnetic grain, two critical parameters that can be tuned to maximize SLP. K$_u$ and V together determine the temperature and frequency dependent coercivity ($H_c$), and the shape of the AC hysteresis loop. Generally, at optimized effective anisotropy, a large NP size is preferred since $M_S$ increases slightly with increasing size, as long as it is not too large to accommodate multi-domain states which reduce $H_c$. Practically, the NPs should be kept nearly SPM to avoid dipole interaction induced agglomeration. We therefore adopt the following strategy to maximize SLP at clinically relevant field parameters: 1. Use soft ferrite Mn$_{0.3}$Fe$_{2.7}$O$_4$ as a starting material to find the largest size for high SLP without agglomeration, as magnetostatic interactions are dependent on $M_S$V ; 2. With size optimized, tune the effective anisotropy to control the barrier of magnetization reversal, by alloying of magnetically hard cobalt ferrite with soft manganese ferrite to maximize SLP. Since MnFe$_2$O$_4$ has a magnetocrystalline anisotropy constant of 3.0×10$^3$ J/m$^3$, while that of CoFe$_2$O$_4$ is 2.0×10$^5$ J/m$^3$, the anisotropy of the mixed ferrite will be very sensitive to the amount of cobalt alloying. At a field of 33 kA/m and 380 kHz, the SLP is maximized by tuning the alloy composition of $Co_xMn_{0.3-x}Fe_{2.7}O_4$. One should aim for a room temperature anisotropy field slightly higher than the field amplitude to achieve the highest SLP.[31] The metal composition is limited to (Co+Mn):Fe = 1:9 to minimize potential toxicity; 3. Maximize SLP and ILP at field parameters below the clinical safety limit ($f$ = 380 kHz, $H \leq$ 13 kA/m and $Hf <$ 5×10$^9$ A/(m·s)). As the anisotropy field of magnetite is larger than the field amplitude, alloying with a soft ferrite such as $Mn_xFe_{3-x}O_4$ is needed. For future *in vivo* applications, however, we choose zinc ferrite due to its bio-compatibility. Stoichiometric $ZnFe_2O_4$ is antiferromagnetic with low magnetization, while non-stoichiometric $Zn_xFe_{3-x}O_4$ is a soft ferrite with $M_S$ moderately higher than that of $Fe_3O_4$ at low Zn content.[25] We thus tune its composition to maximize SLP and ILP at field amplitudes smaller than 13 kA/m.

All types of magnetic cores were synthesized by a modified one-pot thermal decomposition method.[12,34] These NPs are monodisperse with narrow size distribution, as observed from transmission electron microscopy (TEM) images. Fig. 1(a) is the TEM image of 22 nm $Mn_{0.3}Fe_{2.7}O_4$ NPs. NPs show well-defined facets, with polyhedral shape. In our experiments, NPs from 7 nm up to 22 nm were investigated (Fig. S1), below which the NPs show SPM behavior. When the sizes are larger than 22 nm, $Mn_{0.3}Fe_{2.7}O_4$ NPs become ferromagnetic and tend to agglomerate in the solution due to strong magnetostatic interactions. Therefore, they are excluded from further investigations. A silica shell was coated by reverse microemulsion method.[35] The TEM image of the 22-nm $Mn_{0.3}Fe_{2.7}O_4/SiO_2$ NPs is shown in Fig. 1 (b). The silica shells were kept thin with a thickness of 4-5 nm to minimize possible temperature gradient in AC field heating.[22] The silica coating makes the NPs hydrophilic, leading to aqueous dispersions stable for years without agglomeration (Fig. S2(a)). As shown in Fig S2(b), the zeta-potential of the NPs is about -30 mV. Negative charges on NP surface produces sufficient repulsive force to balance the magnetically induced attractive force to keep them from aggregation.

Within the core size range of 7 nm to 22 nm, the SLP of an aqueous dispersion of 1 mg/ml $Mn_{0.3}Fe_{2.7}O_4/SiO_2$ NPs increases monotonically with increasing NP size, under an AC field of 33 kA/m and 380 kHz. The temperature change ΔT *vs* time curves (heating curves) are plotted in Fig. 1 (c), where $\Delta T \equiv T(t)-T_0$; $T(t)$ is the temperature at time t and $T_0$ is ambient temperature. From the heating curves, SLP *vs* size can be extracted using the Box-Lucas fitting known to give reliable SLP values,[36-39] and plotted in Fig. 1(d). The contribution from pure water under identical conditions was also measured and subtracted as the background. As can be seen from Fig. 1(c), the heating curve shows a negligibly small slope for 7 nm NPs, suggesting that 7 nm NPs can hardly heat in such a field; while 10 nm NPs start to heat with a low SLP of 164 $W/g_{metal}$. Towards the other end of the size range, the SLP values for 18-nm NPs are much higher, reaching 1140 $W/g_{metal}$ while 22-nm NPs have the highest value of 2278 $W/g_{metal}$. This can be understood since larger NPs have higher energy barriers for magnetization reversal, thus the area of the AC hysteresis loop will be larger. In Fig. 1(d), the red line is a cubic fitting of SLP as a function of NP diameter *d*. A reasonable agreement between the fitting and experimental data is found, suggesting that SLP is proportional to the NP volume.

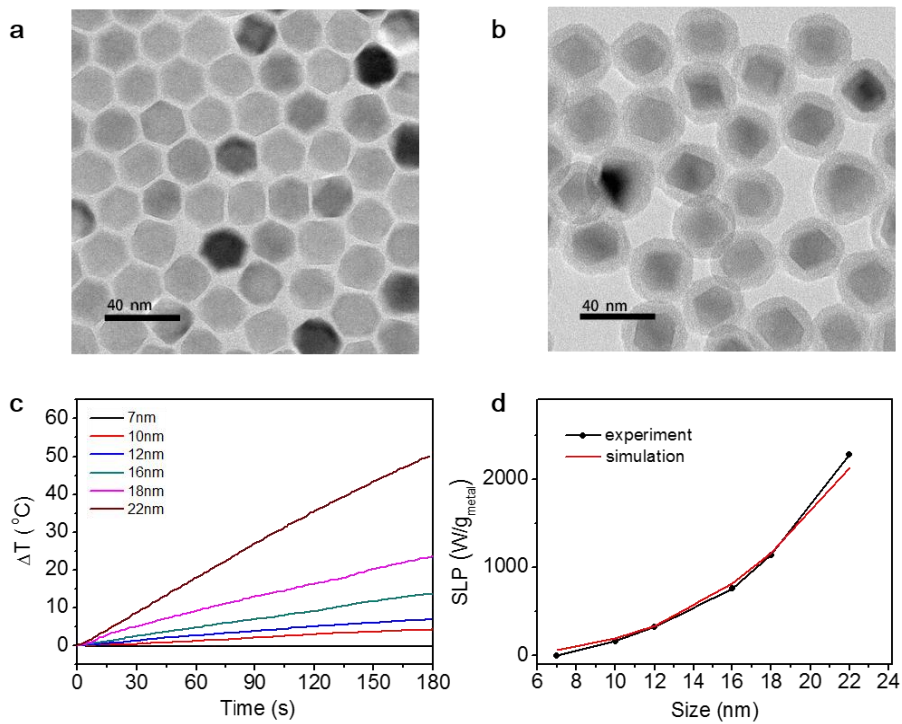

Figure 1. Typical TEM images of (a) 22 nm $Mn_{0.3}Fe_{2.7}O_4$ NPs, (b) 22-nm $Mn_{0.3}Fe_{2.7}O_4/SiO_2$ NPs, (c) Heating curves of aqueous solutions of $Mn_{0.3}Fe_{2.7}O_4/SiO_2$ NPs (1 $mg_{NPs}$/ml) with core sizes from 7 nm to 22 nm; and(d) Size dependence of SLP under the AC field of 380 kHz, 33 kA/m.

Since there is an upper size limit due to agglomeration, as discussed earlier, an alternative approach to enhancing SLP is to increase the effective magnetic anisotropy, while fixing the NP size to 22 nm. We realize anisotropy tuning by cobalt alloying to produce mixed ferrites $Co_xMn_{(0.3-x)}Fe_{2.7}O_4$. Compared to earlier reported core/shell approach,[27] cation alloying possesses distinct advantages: one pot synthesis with high reproducibility, and the ability to tune anisotropy parameters independent of size.

It is found that a small percentage of Co alloying (*i.e.* a moderately larger anisotropy) is optimum for maximizing SLP. The alloy composition (x~0-0.08) was controlled by varying the ratio of Co to Mn precursors. For example, $Co_{0.01}Mn_{0.29}Fe_{2.7}O_4$ NPs were synthesized by mixing Co, Mn and Fe precursors with the molar ratio of 0.01:0.29:2.7. The atomic ratio of Co:Mn:Fe measured by inductively coupled plasma atomic emission spectroscopy (ICP-AES) is 0.27:8.2:91, indicating that the composition of the final product replicates the precursor ratio. The magnetic hysteresis loops of $Co_xMn_{(0.3-x)}Fe_{2.7}O_4$ NPs (x~0-0.03) measured at 300 K and 10 K are shown in Figure 2 (a) to (c). All NPs are SPM at 300 K, while exhibiting hysteresis at 10 K. The 10 K coercivity increases rapidly with increasing Co content, from 27.9 kA/m at x= 0.01 to 46.6 kA/m at x= 0.03. With Co alloying (x~0-0.03), the $M_S$ barely changes. The $M_S$ of $Mn_{0.3}Fe_{2.7}O_4$ NPs is 410 kA/m at 300 K, comparable to the bulk value of 446 kA/m, suggesting excellent crystallinity of the NPs. As shown in Fig. 2(d), the anisotropy constant $K_u$, estimated from low temperature coercivity $H_C$ and $M_S$, increases monotonically with increasing Co concentration. With just 1 at% of Mn substituted by Co (x = 0.03), $K_u$ increases by 300% from $9\times10^3$ J/m³ to $2.8\times10^4$ J/m³. The SLP (Fig. 2(f)), extracted from the heating curves (Fig. 2(e)) measured at $H$ = 33 kA/m and $f$ = 380 kHz, first increases with increasing Co concentration, showing a peak at x=0.03, then decreases with further increasing the Co amount. This is understood since increasing effective anisotropy increases the energy barrier for magnetization reversal, and thus the area of the AC hysteresis loop initially increases. However, with further increasing anisotropy, the field amplitude is insufficient to saturate the magnetization at the operating frequency, resulting in minor hysteresis loops with decreased area. The maximum SLP is measured to be 3417 W/g$_{metal}$ for x= 0.03, which is more than 50% higher than that for the same sized Mn$_{0.3}$Fe$_{2.7}$O$_4$. This value is close to the theoretical limit of SLP for ferrite NPs, with an $\alpha$ value of 0.49. This is higher than $\alpha$ = 0.39 for a random ensemble. We suggest that the higher $\alpha$ not only reflects a certain degree of alignment due to dipole interactions, but also results from an effective anisotropy type close to cubic instead of uniaxial ($M_r/M_S$ = 0.62 and 0.7, for x = 0.02 and 0.03, respectively, as extracted from Fig. 2). We further note that for the sample with maximized SLP, the estimated room temperature anisotropy field $H_K$ is ~ 36 kA/m (see Table 1 of Supporting Information, SI), slightly higher than the applied field of 33 kA/m.

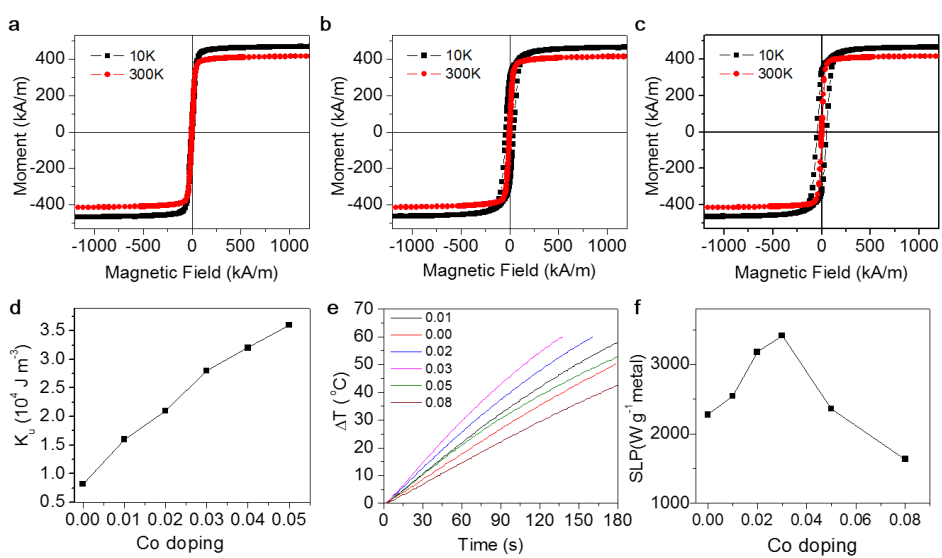

Figure 2. The magnetic hysteresis loops of as-synthesized 22 nm Co$_x$Mn$_{(0.3-x)}$Fe$_{2.7}$O$_4$ NPs with (a) x=0.00, (b) x=0.02, (c) x=0.03, measured at 10 K and 300 K, respectively. (d) Anisotropy constant $K_u$ at 10 K for Co$_x$Mn$_{(0.3-x)}$Fe$_{2.7}$O$_4$ NPs as a function of x; (e) Heating curves of aqueous solutions of Co$_x$Mn$_{(0.3-x)}$Fe$_{2.7}$O$_4$/SiO$_2$ NPs (1 mg$_{NPs}$/ml); and (f) Composition dependence of SLP under the AC field of 380 kHz, 33 kA/m.

In general, SLP values increase with increasing frequency and amplitude of AC fields. However, for clinical hyperthermia applications, there is a safety limit on the product of field frequency and amplitude typically taken to be $Hf$ =5×10$^9$ A/(m·s).[40-42]

Anisotropy optimized to achieve maximum SLP at high fields will not be optimal for clinically relevant low fields. Fig. S5(a) shows the heating curves of 22-nm $Co_{0.03}Mn_{0.27}Fe_{2.7}O_4/SiO_2$ NP aqueous dispersion (1 $mg_{NPs}$/ml) under the AC field of 380 kHz with different field amplitudes. As calculated from the heating curve shown in Fig. S5(a) (see SI), the SLP at $H$ =13 kA/m decreases to just 266 W/g, and further decreases to 40 W/g at 7 kA/m. To achieve large SLP and ILP at fields smaller than 13 kA/m ($Hf < 5 \times 10^9$ A/(m·s)), a softer material with an anisotropy field comparable to the field amplitude is needed. For clinical applications, we resort to non-stoichiometric zinc ferrite NPs for their biocompatibility and low anisotropy.[24] It is known that the $M_S$ of zinc ferrite is highly sensitive to Zn content, with $ZnFe_2O_4$ being antiferromagnetic if $Zn^{2+}$ ions occupy the A-site of the spinel lattice only.[43] One can tune the $M_S$ and anisotropy by varying the Zn: Fe ratio. Fig. 3(a) shows the composition dependence of room temperature $M_S$ in $Zn_xFe_{3-x}O_4$ NPs. It can be seen that $M_S$ increases first monotonically with increasing x, and reaches a maximum of 458 kA/m at x =0.3, comparable to zinc ferrite synthesized by other methods.[44-46] Fig. 3(b) shows the composition dependence of anisotropy $K_u$ measured at 10 K, being nearly constant below x=0.3 and decreases with further increasing x. The room temperature $H_K$ is estimated to be 18.5 kA/m (see SI), suggesting an AC $H_C$ of 9.2 kA/m, which is close to the optimal value for the applied field $H$= 13 kA/m, leading to nearly maximized AC hysteresis area and thus SLP. The heating performance of 22 nm $Zn_{0.3}Fe_{2.7}O_4/SiO_2$ NPs at varying field amplitudes were studied in detail. Fig. S5(b) shows heating curves of $Zn_{0.3}F_{2.7}O_4/SiO_2$ in the AC field of 380 kHz with varying amplitudes, and SLP as a function of field is plotted in Fig. 3(c). Comparing to $Co_{0.03}Mn_{0.27}Fe_{2.7}O_4/SiO_2$, $Zn_{0.3}Fe_{2.7}O_4/SiO_2$ NPs exhibit much higher SLP at fields lower than 13 kA/m ($Hf$ =4.9×10$^9$ A/(m·s), within the clinical safety limit). SLP is 1010 W/$g_{metal}$ at $H$ = 13 kA/m. Achieving SLP > 1000 W/g at clinically safe field parameters is significant, since it would allow sufficient heating for targeted treatment of small tumors and metastases, at low NP concentration achievable through antibody targeting.[47] To compare with other NPs, ILP is calculated and plotted in Fig. 3(d). ILP is 15.7 nHm$^2$/kg at $H$ = 13 kA/m, a value significantly higher than that of NPs synthesized previously.[48] As can be seen from fig. 3(d), ILP of $Zn_{0.3}Fe_{2.7}O_4$ NPs reaches 26.8 nHm$^2$/kg at 7 kA/m (SLP = 500 W/g$_{metal}$). The α values are 0.37 and 0.34 for $H$ = 13 and 7 kA/m, respectively, close to the theoretical value of 0.39 for a completely random ensemble. It is clear that SLP does not scale with $H^2$ as predicted by the linear response theory or ILP would be a constant. As a comparison, bacteria magnetosomes are reported to have a SLP of 960 W/g$_{NPs}$ and ILP of 23.4 nHm$^2$/kg at 410 kHz and 10 kA/m.[32] However, one should note that these values were obtained by aligning the NPs in an external field. Without field alignment, the obtained SLP is expected to be lower by a factor of three.[32] While the SLP increases monotonically with increasing field for both samples as seen in Fig. 3(c), the field dependence shows different behavior. $Co_{0.03}Mn_{0.27}Fe_{2.7}O_4$ NPs have very low initial SLP which increases slowly with field at $H$ < 13 kA/m. This is because the field amplitude of 13 kA/m is insufficient to saturate the magnetization of $Co_{0.03}Mn_{0.27}Fe_{2.7}O_4$ NPs due to their relatively large anisotropy field. The AC hysteresis is expected to be nearly linear in field with very small opening. On the other hand, SLP increases initially rapidly below 18 kA/m then slowly for $Zn_{0.3}Fe_{2.7}O_4$, as its anisotropy field is found to be 18.5 kA/m.

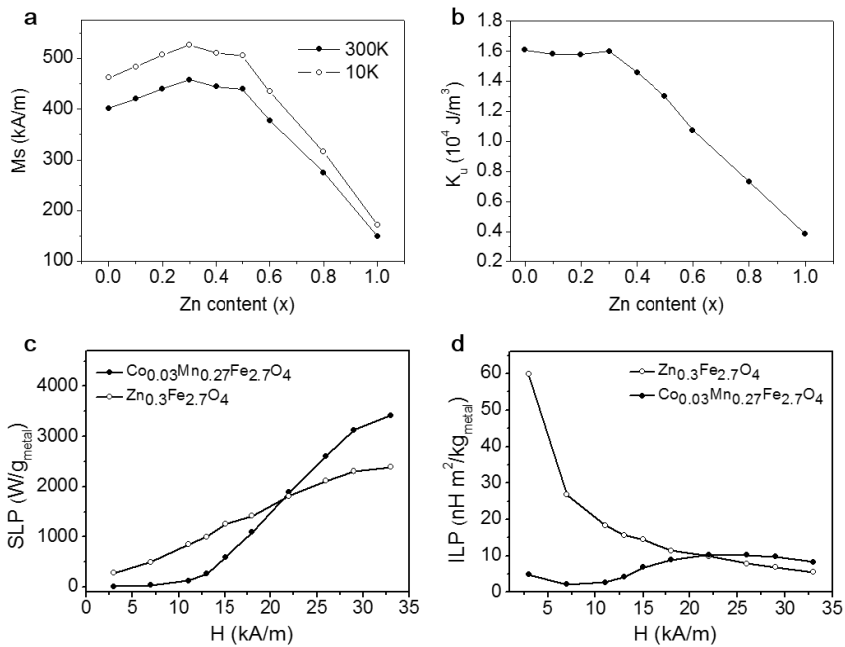

Figure 3. Composition dependence of (a) saturation magnetization $M_S$, measured at 10 and 300 K; and (b) anisotropy constant $K_u$ at 10 K, for $Zn_xFe_{3-x}O_4$ NPs. (c) Field dependence of SLP for $Zn_{0.3}Fe_{2.7}O_4/SiO_2$ and $Co_{0.03}Mn_{0.27}Fe_{2.7}O_4/SiO_2$ NPs, (d) Field dependence of ILP for $Zn_{0.3}Fe_{2.7}O_4/SiO_2$ and $Co_{0.03}Mn_{0.27}Fe_{2.7}O_4/SiO_2$ NPs.

In clinical hyperthermia applications, the contribution to heating due to physical rotation of NPs (Brownian relaxation) may be hindered in biological environment, e.g. in NPs embedded in bone cement. To investigate the realistic heating performance, we studied AC field heating of NP dispersions in water/glycerol mixture with different glycerol concentration up to 80 vol%. As can be seen in Fig. S10, SLP decreases with increasing glycerol concentration at all field values studied. At 80 vol% glycerol, the SLP ranges from 50% (at 3 kA/m) to 70% (at 18 kA/m) of the values for aqueous solutions. The reduction is primarily due to the high viscosity of glycerol, which is 60 times that of water at room temperature.[49] The high viscosity of the glycerol solution hinders the rotation of the NPs, making Brownian motion ineffective in contributing to AC field heating. However, it is clear that the dominating contribution to hysteresis loss is the Néel relaxation, as more than 50% of the heating performance is retained for the most viscos sample. This enables high heating of NP bone cement as discussed in the following section.

Surgical resection combined with chemo- and radiotherapy has been a clinical gold standard for the treatment of bone tumors. However, the patients are more likely to experience a tumor recurrence due to the bone microenvironment-associated tumor resistance to chemo- and radiotherapy, or inadequate surgical margins. Targeted thermal therapy of bone tumors become attractive because of its high selective damage of tumor tissue and repeatability.[50] We use bone cement containing 1 wt% $Zn_{0.3}Fe_{2.7}O_4$ NPs for local hyperthermia experiments. A piece of pig rib with a hole 6.0 mm in diameter and 6.0 mm in length were filled with the magnetic cement, and exposed to an AC field of 380 kHz and 13 kA/m, as shown in Fig. 4 (a). The surrounding of the bone was water-cooled at 37 °C to mimic cooling by blood vessels. The temperature rise of the pig rib was recorded by both a high-resolution infrared (IR) camera and a fiber optic probe. It can be seen that the temperature of the magnetic cement rises rapidly to the therapeutic threshold required for cancer hyperthermia (T >42 ℃). The center of cement reaches 50 ℃ within 1 minute; and the temperature of entire cement rises to above 50 ℃ within 3 min. When the heating time is up to 30 min, the cement is heated to 70 ℃, while the region 25 mm away from the center is above 46 ℃. In Fig. 4 (a), the red dots represent the region with the temperature at the therapeutic threshold of 42 ℃. With increasing exposure time, the region with temperature > 42 ℃ expands. After 25 min, the temperature of the entire bone is over 42 ℃. As a comparison, in a previous study,[51] bone cement containing 60 wt% of magnetic materials was used to reach similar temperature change at a maximum AC field of 100 kHz and 23.9 kA/m. A significant reduction in dosage afforded by the high SLP of our optimized bio-compatible NPs can not only minimize any potential toxicity, but also preserve the mechanical integrity of the bone cement.

Two types of NPs, 22-nm $Co_{0.03}Mn_{0.27}Fe_{2.7}O_4/SiO_2$ and $Zn_{0.3}Fe_{2.7}O_4/SiO_2$, were tested for hyperthermia killing of Osteosarcoma MG-63 cells. Cell apoptosis was examined by a flow-cytometry-based annexin-V fluorescein isothiocyanate (see SI), as shown in Fig. 4 (b) to (e), respectively. Cells without NPs and with NPs but no AC field exposure were used as control. $1 \times 10^4$ MG-63 cells and $Co_{0.03}Mn_{0.27}Fe_{2.7}O_4/SiO_2$ NP solution with a concentration of 25 μg/ml were exposed to an AC magnetic field of 380 kHz, 33 kA/m for 3 min. The percentage of the early and late apoptotic cells (*i.e.* cell death) is 81% in total, similar to values reported previously using a dosage of 50 $μg_{NPs}$/ml at 37.4 kA/m and 500 kHz ($Hf$ value 1.5 times higher).[16] To test the performance of $Zn_{0.3}Fe_{2.7}O_4/SiO_2$ NPs, a concentration of 300 $μg_{NPs}$/ml were used. Exposure to an AC field of $H$ = 13 kA/m for 10 min resulted in 89% cell death, where the early apoptotic cells was 79.35% and late apoptotic cells 9.83%. It should be emphasized that all hyperthermia treatments in our studies were performed on adherent cells. Hyperthermia on suspended cells may overestimate cell death, as cells would be inevitably injured during the digestive process.[18,39,52] Since adherent cells more closely simulate cells in vivo than suspended cells, our reported values should be more relevant in guiding clinical applications.

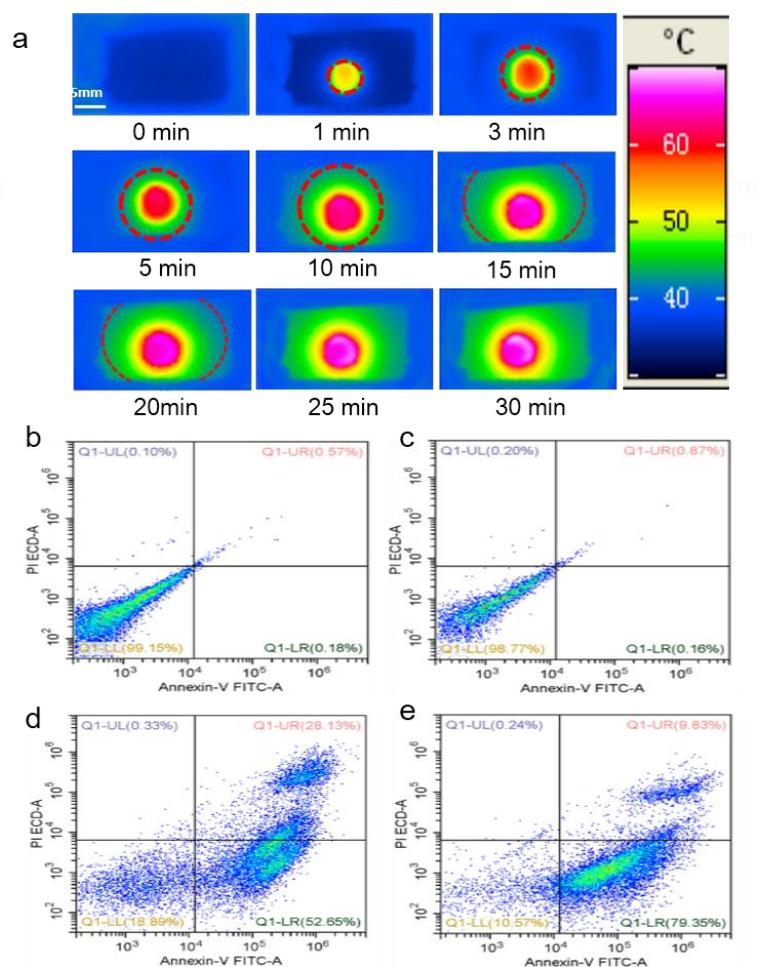

Figure 4. (a) Infrared Photos of the bone heated under the field of 380 kHz, 13 kA/m; Apoptosis assay fluorescence from Annexin V and PI uptake by the MG-63 cells were monitored by flow cytometry. (b) MG-63 cells without NPs used as Control group. (c) MG-63 cells with NPs but not heated used as the second Control group. (d) MG-63 cells incubated with $25\mu g_{NPs}$/ml $Co_{0.03}Mn_{0.27}Fe_{2.7}O_4$/$SiO_2$ NPs heated under the AC field of 380 kHz, 33 kA/m (e) MG-63 cells incubated with 300 $\mu g_{NPs}$/ml $Zn_{0.3}Fe_{2.7}O_4$/$SiO_2$NPs heated under the AC field of 380 kHz, 13 kA/m.

Cytotoxicity of different types of NPs to cells was also measured and compared. The viability of the MEF and MG-63 cells was determined by CCK-8 assay after incubation with various concentrations of NPs ($Zn_{0.3}Fe_{2.7}O_4$/$SiO_2$, $Fe_3O_4$/$SiO_2$, and $Mn_{0.3}Fe_{2.7}O_4$/$SiO_2$) for 24 and 48h. Cells without NPs were used as control groups. As shown in Fig. 5, the cytotoxicity is the lowest for $Zn_{0.3}Fe_{2.7}O_4$/$SiO_2$ and highest for $Co_{0.03}Mn_{0.27}Fe_{2.7}O_4$/$SiO_2$, with $Fe_3O_4$/$SiO_2$ in between. In the case of $Zn_{0.3}Fe_{2.7}O_4$/$SiO_2$, up to the concentration of 1000 μg/ml, the cell viability shows no significant difference (P > 0.05) from the control groups after incubation for 24h. Though at 700 and 1000 μg/ml after incubation for 48 h, viability of cells with $Zn_{0.3}Fe_{2.7}O_4/SiO_2$ is lower than that of control groups, it is still above 75% for MEF and above 73% for MG-63 cells. For MEF Incubated with $Fe_3O_4/SiO_2$ NPs at concentration higher than 500 μg/ml at both incubation time of 24 and 48 h, cell viability is lower than that of the control sample, while $Fe_3O_4/SiO_2$ shows cytotoxicity in MG-63 starting at 300 μg/ml upon incubation for 48 h. As for $Co_{0.03}Mn_{0.27}Fe_{2.7}O_4/SiO_2$, cytotoxicity is observed at concentration above 100 μg/ml, regardless of cell lines and incubation time. At the highest concentration of 1000 μg/ml and 48 h, cell viability is even as low as 9.48%.

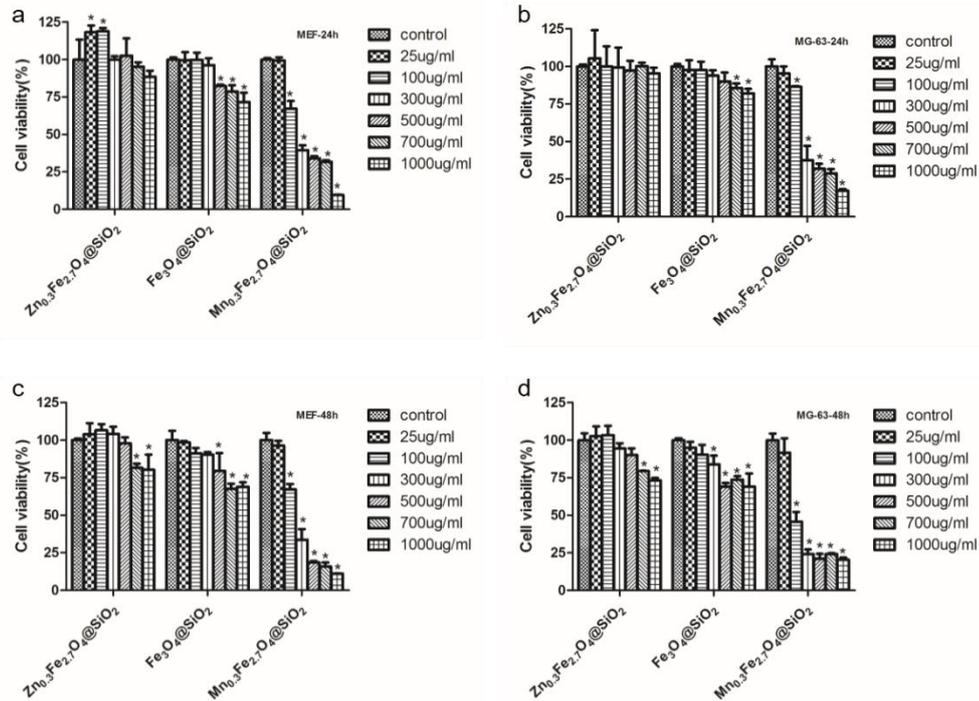

Figure 5. The viability of the MEF and MG-63 cells determined by CCK-8 assay after incubation in NP solutions with various concentrations for 24h (a,b) and 48h (c,d).

In summary, we have designed and synthesized two types of magnetic/silica core/shell NPs. $Co_xMn_{0.3-x}Fe_{2.7}O_4/SiO_2$ with Co concentration of x=0.03 results in maximized specific loss power of 3417 W/g at an AC field of 33 kA/m and 380 kHz; and biocompatible $Zn_{0.3}Fe_{2.7}O_4/SiO_2$ achieved SLP of 1010 W/g at a field of 13 kA/m and 380 kHz. The intrinsic loss power ranges from 15.7 to 59.9 nHm$^2$/kg as the field decreases from 13 to 3 kA/m for the latter. We further demonstrate efficient hyperthermia using Zn$_{0.3}$Fe$_{2.7}$O$_4$ NPs in magnetic cement for bone tumor, incorporating ultralow dosage of just 1 wt% of nanoparticles. Zn$_{0.3}$Fe$_{2.7}$O$_4$ NPs also demonstrate good hyperthermia performance to kill cancer cells. Zn$_{0.3}$Fe$_{2.7}$O$_4$ NPs show excellent biocompatibility, exhibiting no cell cytotoxicity at concentrations up to 500 μg/ml within 48 hrs. Our work provides a guidance for design of NPs with appropriate magnetic properties for maximized heating power at any field parameters, and conversely, given a particular NP type, choice of field parameters leading to maximized heating power. Furthermore, our biocompatible NP platform with greatly enhanced AC field heating at low field amplitudes are promising for targeted hyperthermia of small tumors and metastases. Further in vivo studies are needed to show the therapeutic effect of these optimized nanoparticles.

**Methods**

1. **Experimental Methods**

   **Synthesis of $Co_xMn_{(1-x)}Fe_2O_4$ nanoparticles**

   A series of $Co_xMn_{(0.3-x)}Fe_{2.7}O_4$ mixed ferrite NPs of different sizes and composition were synthesized by a one-pot solution method through thermal decomposition of a mixture of metal acetylacetonates with surfactants in a high-boiling point organic solvent. Iron(III) acetylacetonate (2 mmol), manganese(II) acetylacetonate and cobalt(II) acetylacetonate (total 1 mmol), 1,2-hexadecanediol, oleic acid (3 mmol), oleylamine (3 mmol), and 20 mL benzyl ether were mixed and magnetically stirred under a flow of nitrogen. The mixture was first heated to 393 K for 30 minutes to remove the low boiling point solvent, then to 473 K and kept at that temperature for 1 hour. At a ramping rate of 10 K min$^{-1}$ the solution was further heated to reflux (~573 K) and kept at 573 K for 1 hour. The solution was cooled down to room temperature by removing the heat source. The solution was treated with ethanol in the air atmosphere. $Co_xMn_{(1-x)}Fe_2O_4$ NPs were precipitated from the solution, centrifuged to remove the solvent, and redispersed in hexane. The size of nanoparticles was tuned by the ratio of oleic acid to metal precursors.[12]

   **Synthesis of $Zn_{0.3}Fe_{2.7}O_4$ nanoparticles**

   Under a gentle flow of Ar, Iron(III) acetylacetonate (2.7 mmol), zinc(II) acetylacetonate (0.3 mmol), sodium oleate (2 mmol) and oleic acid (4 ml) were mixed with benzyl ether (20 ml). The mixture was magnetically stirred under a flow of Ar and then heated to 393 K for 1 h. Under an Ar blanket, the solution was further heated to reflux (~573 K) and kept at this temperature for 1h. The mixture was then cooled down to room temperature by removing the heating mantle. The size of nanoparticles were tuned by controlling the heating rate during heating from 393 K to 573 K.

   **Silica coating of magnetic NPs**

   The silica shells were coated on the hydrophobic NPs *via* a reverse microemulsion method.[22,32] For 4-5 nm silica coating, 20 ml cyclohexane and 1.15 ml Igepal CO-520 were mixed and 20 mg magnetic NPs in 2 ml cyclohexane were added while stirring. 0.15 ml ammonium hydroxide (28-30%) was then added, followed by 0.1 ml TEOS. The solution was stirred at room temperature for 24 h and the resulting magnetic NPs/$SiO_2$ core/shell NPs were precipitated by adding ethanol and centrifugation. The collected particles were washed in ethanol and water twice and precipitated by centrifugation and finally redispersed in water.

### Characterizations

A Hitachi H7650 (120kV) transmission electron microscope was used to characterize the size and morphology of the NPs. Energy dispersive X-ray spectroscopy and inductively coupled plasma atomic emission spectroscopy were employed to determine the composition. The magnetic hysteresis loops were measured using a Quantum Design Physical Property Measurement System model 6000. Hyperthermia performance of the NPs was investigated by a HYPER5 machine fabricated by MSI Company under an AC magnetic field with frequency of 380 kHz. The temperature change of the NP solution was monitored by a fiber optic probe.

### In vitro experiments

Human osteogenic sarcoma MG-63 cells and mouse fibroblast cells (MEF) were purchased from the American Type Culture Collection. MG-63 cells were plated in 35-mm culture dishes at 80% confluence ($1 \times 10^6$ cells) with 2ml of DMEM/HIGH GLUCOSE medium containing 10% fetal bovine serum. The NP dispersion was added to culture dishes, and the samples were exposed to the AC magnetic field. Apoptotic cell was detected by a flow cytometer (Beckman coulter Ltd., USA). MEF cells were seeded in 96-well plates at a density of 5,000 cells per well. After incubation for 24 hr and 48 hr, the cell viabilities were determined by the standard Cell Counting Kit-8 (CCK-8, Dojindo, Japan) assay.

**Acknowledgements**

This work was supported by National Science Foundation of China (Grant No. 51571146, 51471186, 51372276).

# Supporting Information

**Maximizing Specific Loss Power for Magnetic Hyperthermia by Hard-Soft Mixed Ferrites**

Shuli He[1,2,4], Hongwang Zhang[2], Yihao Liu[1,3], Fan Sun[2], Xiang Yu[1], Xueyan Li[1], Li Zhang[1], Lichen Wang[1], Keya Mao[3], Gangshi Wang[3], Yunjuan Lin[3], Zhenchuan Han[3], Renat Sabirianov[5], and Hao Zeng[2*]

[1] Department of Physics, Capital Normal University, Beijing 100048, China

[2] Department of Physics, University at Buffalo, SUNY, Buffalo, New York 14260, USA

[3] Chinese PLA General Hospital, Beijing 100853, China

[4] Beijing Advanced Innovation Center for Imaging Technology, Beijing 100048, China

[5] Department of Physics, University of Nebraska-Omaha, Omaha, NE 68182, USA

*corresponding author: haozeng@buffalo.edu

## 1. Experimental Methods

Synthesis of Co$_x$Mn$_{(0.3-x)}$Fe$_2$O$_4$ nanoparticles

A series of Co$_x$Mn$_{(0.3-x)}$Fe$_2$O$_4$ mixed ferrite NPs of different sizes and composition were synthesized by a one-pot solution method through thermal decomposition of a mixture of metal acetylacetonates with surfactants in a high-boiling point organic solvent. Iron(III) acetylacetonate (2 mmol), manganese(II) acetylacetonate and cobalt(II) acetylacetonate (total 1 mmol), 1,2-hexadecanediol, oleic acid (3 mmol), oleylamine (3 mmol), and 20 mL benzyl ether were mixed and magnetically stirred under a flow of nitrogen. The mixture was first heated to 393 K for 30 minutes to remove the low boiling point solvent, then to 473 K and kept at that temperature for 1 hour. At a ramping rate of 10 K min$^{-1}$ the solution was further heated to reflux (~573 K) and kept at 573 K for 1 hour. The solution was cooled down to room temperature by removing the heat source. The solution was treated with ethanol in the air atmosphere. Co$_x$Mn$_{(0.3-x)}$Fe$_2$O$_4$ NPs were precipitated from the solution, centrifuged to remove the solvent, and redispersed in hexane. The size of nanoparticles was tuned by the ratio of oleic acid to metal precursors.

Synthesis of Zn$_{0.3}$Fe$_{2.7}$O$_4$ nanoparticles

Under a gentle flow of Ar, Iron(III) acetylacetonate (2.7 mmol), zinc(II) acetylacetonate (0.3 mmol), sodium oleate (2 mmol) and oleic acid (4 ml) were mixed with benzyl ether (20 ml). The mixture was magnetically stirred under a flow of Ar and then heated to 393 K for 1 h. Under an Ar blanket, the solution was further heated to reflux (~573 K) and kept at this temperature for 1h. The mixture was then cooled down to room temperature by removing the heating mantle. The size of nanoparticles were tuned by controlling the heating rate during heating from 393 K to 573 K.

Silica coating of magnetic NPs

The silica shells were coated on the hydrophobic NPs via a reverse microemulsion method. For 4-5 nm silica coating, 20 ml cyclohexane and 1.15 ml Igepal CO-520 were mixed and 20 mg magnetic NPs in 2 ml cyclohexane were added while stirring. 0.15 ml ammonium hydroxide (28-30%) was then added, followed by 0.1 ml TEOS. The solution was stirred at room temperature for 24 h and the resulting magnetic NPs/SiO2 core/shell NPs were precipitated by adding ethanol and centrifugation. The collected particles were washed in ethanol and water twice and precipitated by centrifugation and finally redispersed in water.

Characterizations

A Hitachi H7650 (120kV) transmission electron microscope was used to characterize the size and morphology of the NPs. Energy dispersive X-ray spectroscopy and inductively coupled plasma atomic emission spectroscopy were employed to determine the composition. The magnetic hysteresis loops were measured using a Quantum Design Physical Property Measurement System model 6000. Hyperthermia performance of the NPs was investigated by a HYPER5 machine fabricated by MSI Company under an AC magnetic field with frequency of 380 kHz. The temperature change of the NP solution was monitored by a fiber optic probe.

In vitro experiments

Human osteogenic sarcoma MG-63 cells and mouse fibroblast cells (MEF) were purchased from the American Type Culture Collection. MG-63 cells were plated in 35-mm culture dishes at 80% confluence ($1 \times 10^6$ cells) with 2ml of DMEM/HIGH GLUCOSE medium containing 10% fetal bovine serum. The NP dispersion was added to culture dishes, and the samples were exposed to the AC magnetic field. Apoptotic cell was detected by a flow cytometer (Beckman coulter Ltd., USA). MEF cells were seeded in 96-well plates at a density of 5,000 cells per well. After incubation for 24 hr and 48 hr, the cell viabilities were determined by the standard Cell Counting Kit-8 (CCK-8, Dojindo, Japan) assay.

## 2. TEM Images of Mn$_{0.3}$Fe$_{2.7}$O$_4$ and Mn$_{0.3}$Fe$_{2.7}$O$_4$/SiO$_2$ NPs with different sizes

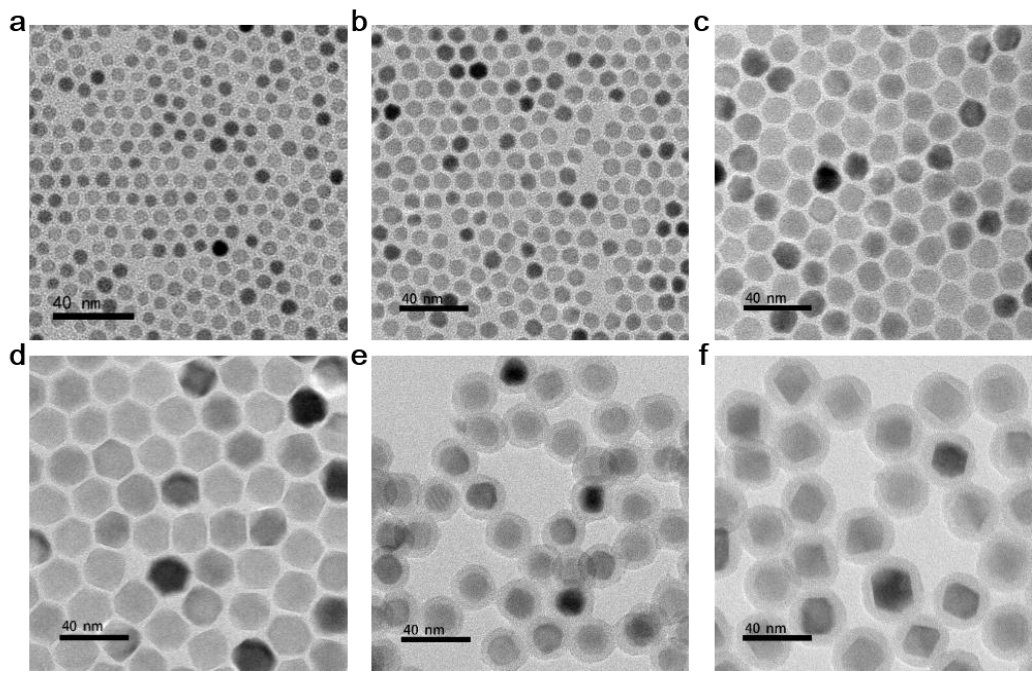

Figure S1. TEM images of Mn$_{0.3}$Fe$_{2.7}$O$_4$ nanoparticles of (a) 7 nm (b) 10nm, (c) 16 nm, (d) 22 nm, (e) 16-nm Mn$_{0.3}$Fe$_{2.7}$O$_4$/5-nm SiO$_2$ NPs, (f) 22-nm Mn$_{0.3}$Fe$_{2.7}$O$_4$/5-nm SiO$_2$ NPs

## 3. Long term stability of the NP dispersion

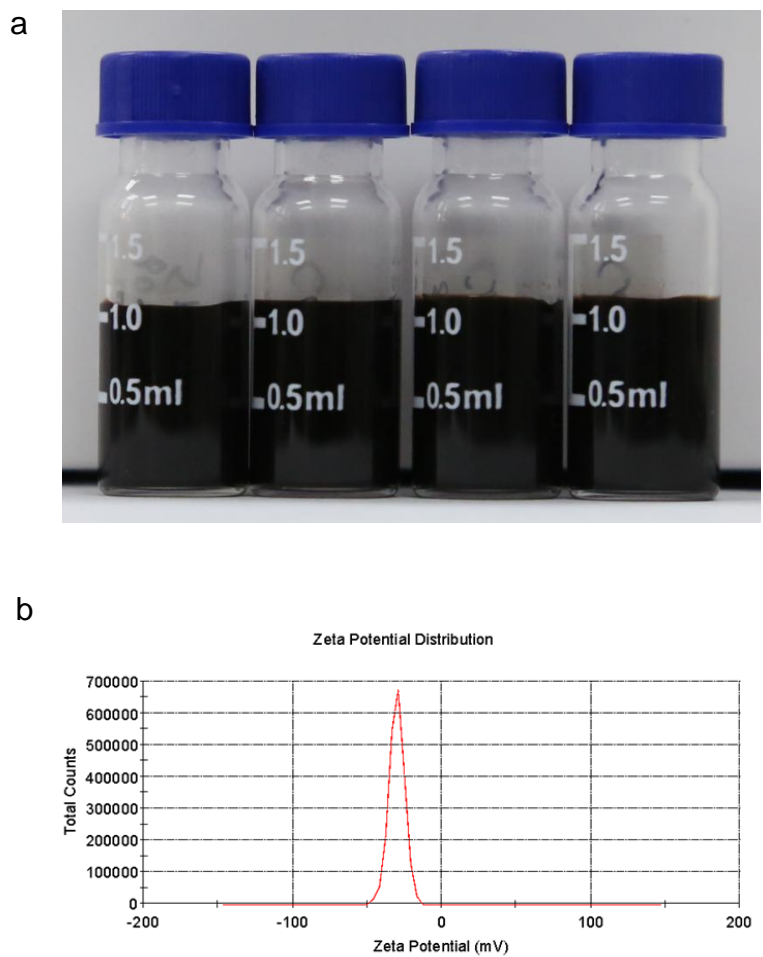

Figure S2. (a) A photograph of Co$_x$Mn$_{0.3-x}$Fe$_{2.7}$O$_4$/SiO$_2$ aqueous dispersions (10 mg$_{NPs}$ $ml^{-1}$), from left to right: x=0.00, x=0.01, x=0.02, x=0.03. The samples are stable for > 24 months with no precipitation, (b) z-potential curve of Co$_{0.03}$Mn$_{0.27}$Fe$_{2.7}$O$_4$/SiO$_2$ nanoparticles.

As shown in Fig S2(b), zeta potential of Co$_x$Mn$_{0.3-x}$Fe$_{2.7}$O$_4$/SiO$_2$ NPs is about -30 mV. Negative charges on NP surface produces sufficient repulsive force to balance the magnetically induced attractive force in water.

## 4. Synthesis of $Zn_{0.3}Fe_{2.7}O_4$ NPs -Tuning of NP sizes by heating rate

The sizes of $Zn_{0.3}Fe_{2.7}O_4$ nanoparticles can be tuned by controlling the heating rate during the synthesis from 393 K to 573 K. The heating rate of 10 K/min leads to 22-nm ZFO NPs, and 6 K/min results in 18-nm ZFO NPs.

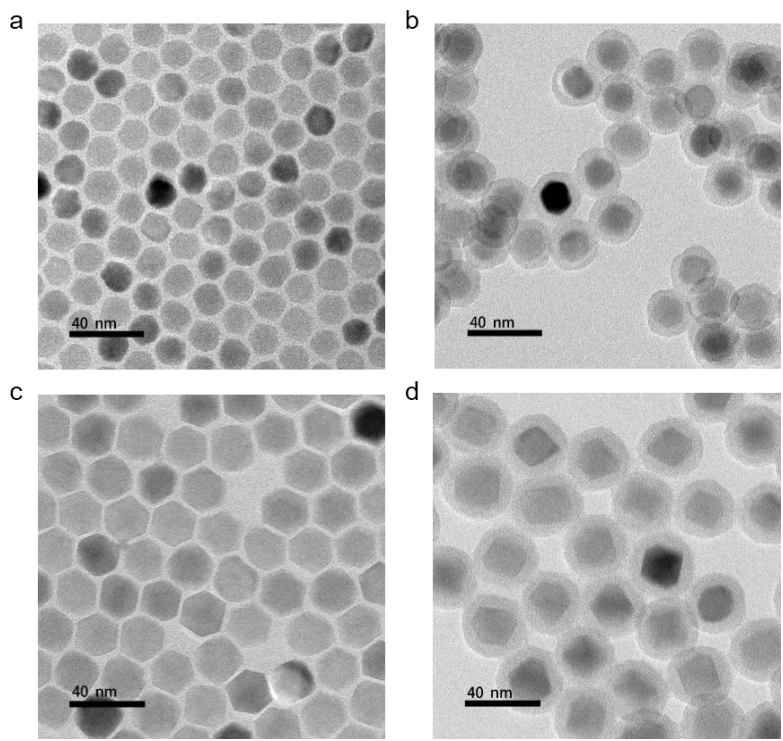

Figure S3. TEM images of (a) $Zn_{0.3}Fe_{2.7}O_4$ NPs, (b) $Zn_{0.3}Fe_{2.7}O_4@SiO_2$ NPs synthesized with the heating rate of 6 K/min, (c) $Zn_{0.3}Fe_{2.7}O_4$ NPs, (d) $Zn_{0.3}Fe_{2.7}O_4@SiO_2$ NPs synthesized with the heating rate of 10K/min.

## 5. Measurement of magnetic anisotropy

Saturation magnetization $M_S$ of NPs was measured by an EV-9 vibrating sample magnetometer. NPs were embedded in non-magnetic cement (purchased from Quantum Design company). The mass of organic coating on NPs was measured by thermogravimetric analysis. The net mass of NPs was used to determine the saturation magnetization. The effective magnetic anisotropy $K_u$ was estimated from $M_S$ and $H_C$ at 10 K, using $K_u \sim M_s H_c$.

Table 1. Saturation magnetization $M_S$ measured at 10 and 300 K, coercitivity $H_C$ measured at 10 K, anisotropy $K_u$ at 10 K, and estimated anisotropy field $H_K$ at 300 K of $Co_xMn_{0.3-x}Fe_{2.7}O_4$ NPs. $K_u$ increases with increasing Co alloying concentration x.

| $Co_xMn_{0.3-x}Fe_{2.7}O_4$ | x =0.00 | x =0.01 | x =0.02 | x =0.03 |
|---|---|---|---|---|
| $M_S$ (kA/m) at 10 K | 468.4 | 466.9 | 465.8 | 465.3 |
| $M_S$ (kA/m) at 300 K | 410.5 | 409.5 | 408.9 | 408.4 |
| $H_C$ (kA/m) at 10 K | 15.0 | 27.9 | 37.2 | 46.6 |
| $K_u$ ($10^4$ J/m$^3$) at 10 K | 0.9 | 1.66 | 2.2 | 2.8 |
| $H_K$ (kA/m) at 300 K | 11.5 | 21.5 | 28.7 | 35.9 |

Table 2. Saturation magnetization $M_S$ measured at 10 and 300 K, coercitivity $H_C$ measured at 10 K, anisotropy $K_u$ at 10 K, and estimated anisotropy field $H_K$ at 300 K of $Zn_xFe_{3-x}O_4$ NPs.

| $Zn_xFe_{3-x}O_4$ | x =0.0 | x =0.2 | x =0.3 | x =0.4 | x =0.5 | x =0.8 | x =1.0 |
|---|---|---|---|---|---|---|---|
| $M_S$ (kA/m) at 10 K | 461.7 | 506.8 | 526.8 | 511 | 505.3 | 316.1 | 171.7 |
| $M_S$ (kA/m) at 300 K | 401.5 | 440.7 | 458.1 | 444.3 | 439.4 | 274.9 | 149.3 |
| $H_C$ (kA/m) at 10 K | 28.2 | 25.3 | 24.6 | 23.1 | 20.8 | 18.7 | 18.1 |
| $K_u$ ($10^4$ J/m$^3$) at 10 K | 1.61 | 1.57 | 1.6 | 1.45 | 1.29 | 0.73 | 0.38 |
| $H_K$ (kA/m) at 300 K | 21.3 | 19.1 | 18.5 | 17.4 | 15.7 | 14.1 | 13.7 |

$H_K$ at 300 K is estimated using the formula below to consider the thermal fluctuation of magnetic moment; γ is taken to be 3 [S1]

$$\frac{K_u(T=0)}{K_u(T=300\ K)} = \left[\frac{M_S(T=0)}{M_S(T=300\ K)}\right]^{\gamma}$$

## 6. Measurement of solution temperature for SLP determination

The temperature of the solution was measured by a fiber optic thermometer at different locations of the vial. A schematic drawing of the measurement setup is shown in Fig. S4 (a). The heating curves for these locations are plotted in Fig. S4 (b).

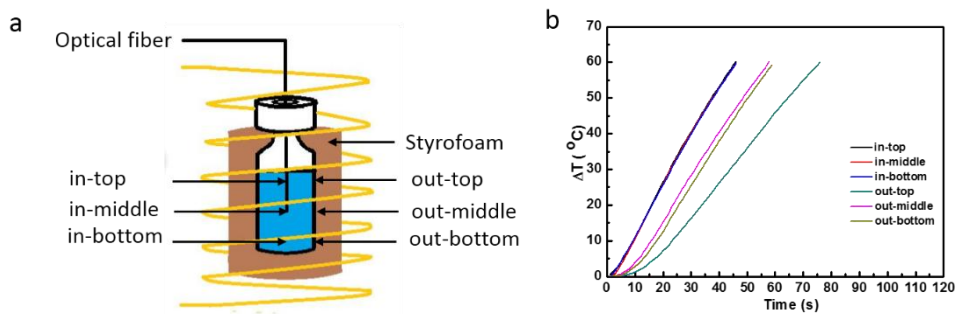

Fig. S4. (a) Schematic drawing of the measurement setup used in this work; (b) The heating curves measured at different locations (including outside the vial).

As can be seen from Fig. S4(b), the temperature inside the vial is uniform, with negligible temperature difference for the probe located at the top, middle and bottom of the solution in the vial. All SLP reported were measured with the probe located at middle of the solution in the vial. Furthermore, the vial temperature, measured at the outer surface of the vial, is considerably higher than the ambient temperature, suggesting that the energy absorption by the container cannot be neglected. As can be seen from the heating curves for the probe located at the outer surface of the vial, the surface temperature change of the vial can reach 2/3 of the solution temperature change. For example, as the solution temperature increases by 60 °C, the vial temperature can increase by 40 °C. Therefore, we stress that measurements ignoring the energy absorption of the container tends to underestimate the SLP values. However, since most previous papers published SLP values without considering container absorption, in this paper all SLP values were calculated ignoring the vial absorption, for a fair comparison with published results.

## 7. AC filed heating curves for two type of NPs at different field amplitudes

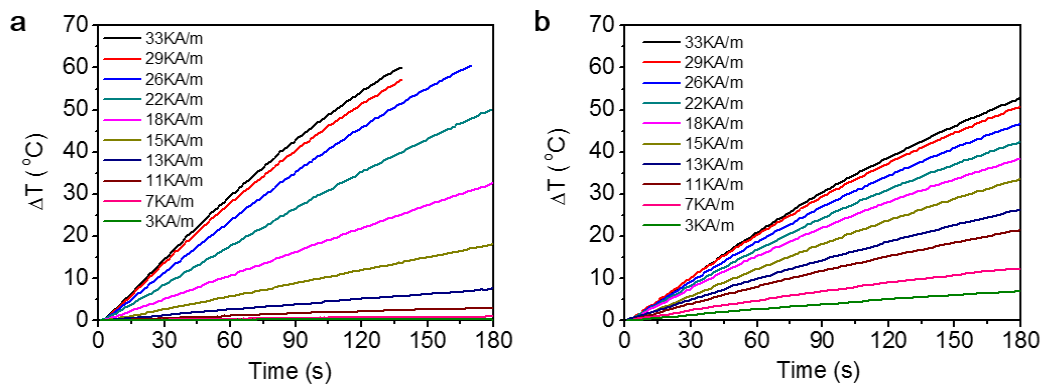

Figure S5. AC field heating of NPs measured by temperature *vs* time curves for (a) $Co_{0.03}Mn_{0.27}Fe_{2.7}O_4/SiO_2$ NP and (b) $Zn_{0.3}Fe_{2.7}O_4/SiO_2$ NP aqueous dispersions (1 mg$_{NPs}$/ml) under an AC field of 380 kHz with different field amplitudes.

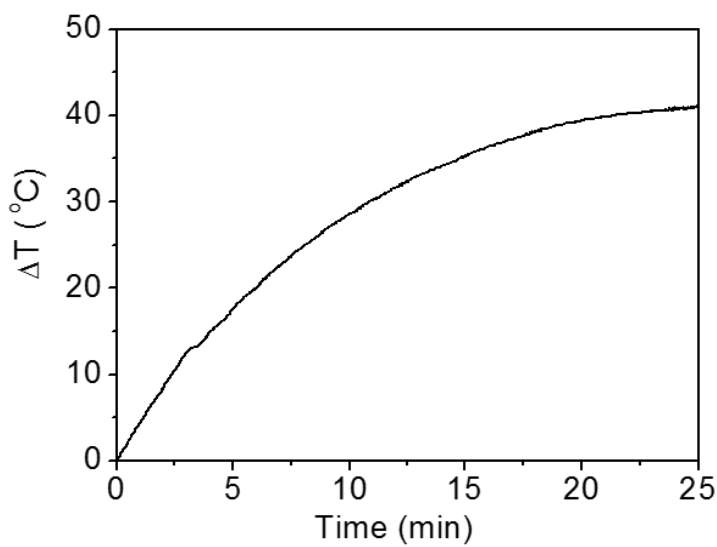

Figure S6. AC field heating curve for $Zn_{0.3}Fe_{2.7}O_4/SiO_2$ NP aqueous dispersions (5 mg$_{NPs}$/ml) under an AC field of 3 kA/m and 380 kHz. The temperature change reaches 40 °C in 25 min, demonstrating the remarkable heating capability of this material at ultralow field.

## 8. Extraction of SLP

For reliable extraction of SLP, all heating curves are fitted by the Box-Lucas formula $\Delta T = \frac{S_m}{k}(1 - e^{-k(t-t_0)})$[S2] with $S_m$ and $k$ as the fitting parameters. $S_m$ is the initial slope of the heating curve, and $k$ is a constant describing the cooling rate. SLP is then calculated as $SLP = \frac{C_v S_m}{\rho_i}$, where $C_v$ is the specific heat capacity of the solution taken to be 4.184 J/(g·°C), and $\rho_i$ is the mass concentration of the metal in the NP solution (*e.g.* for $Fe_3O_4$, 1 $mg_{NPs}$/ml = 0.724 $mg_{Fe}$/ml).

The reliability of the fitting is further verified by measuring the cooling curve directly (shown in Fig. S7(b)), from which $k$ can be extracted independently using $\Delta T = T_0 + (T - T_0)e^{-k(t-t_0)}$. The difference is found to be smaller than 0.1%.

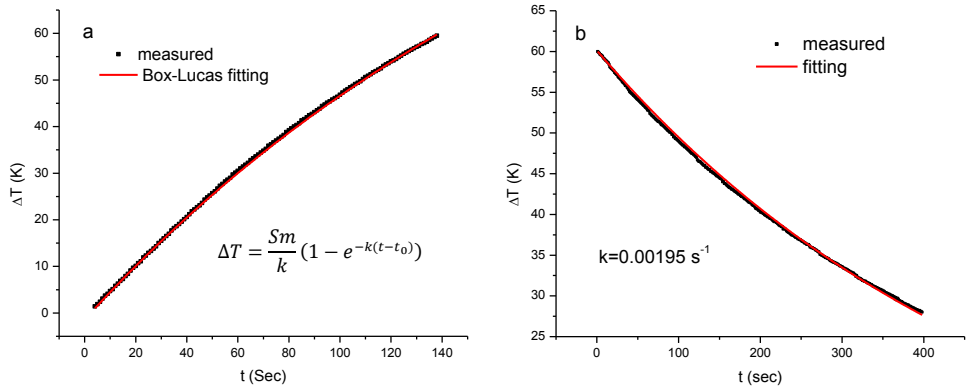

Figure S7. (a) Experimental heating curve (black) and Box-Lucas fitting (red) of $Co_{0.03}Mn_{0.27}Fe_{2.7}O_4$/$SiO_2$ NP solution exposed in the ac field of 380 kHz and 33 kA/m, (b) Experimental cooling curve of NPs solution (black) and fitting curve using $T(t) = T_0 + (T - T_0)e^{-k(t-t_0)}$ (red).

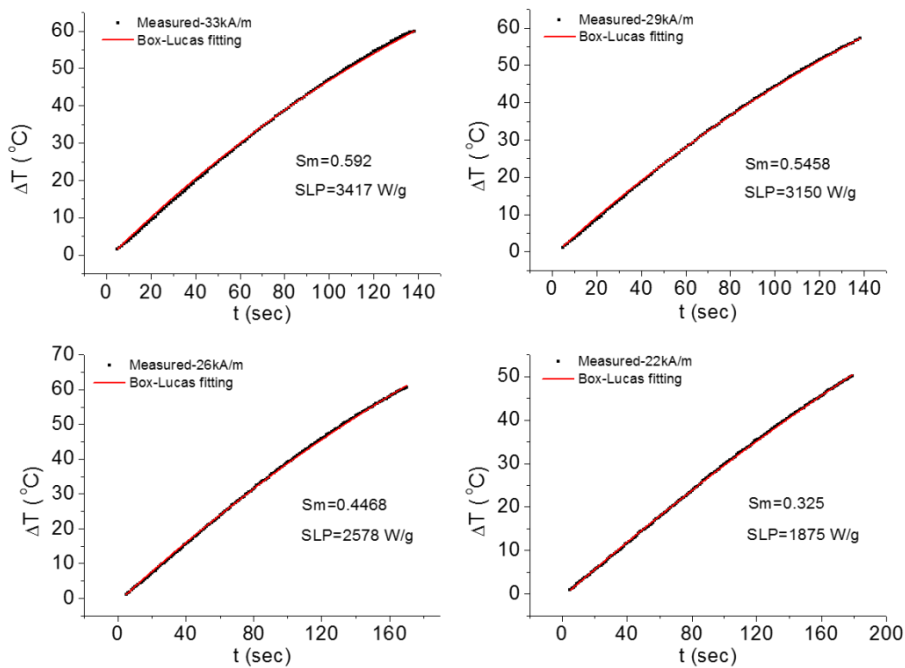

Figure S8. Examples of heating curve (black) and Box-Lucas fitting curve (red) of $Co_{0.03}Mn_{0.27}Fe_{2.7}O_4/SiO_2$ NP solution exposed in the AC field of 380 kHz under different field amplitudes.

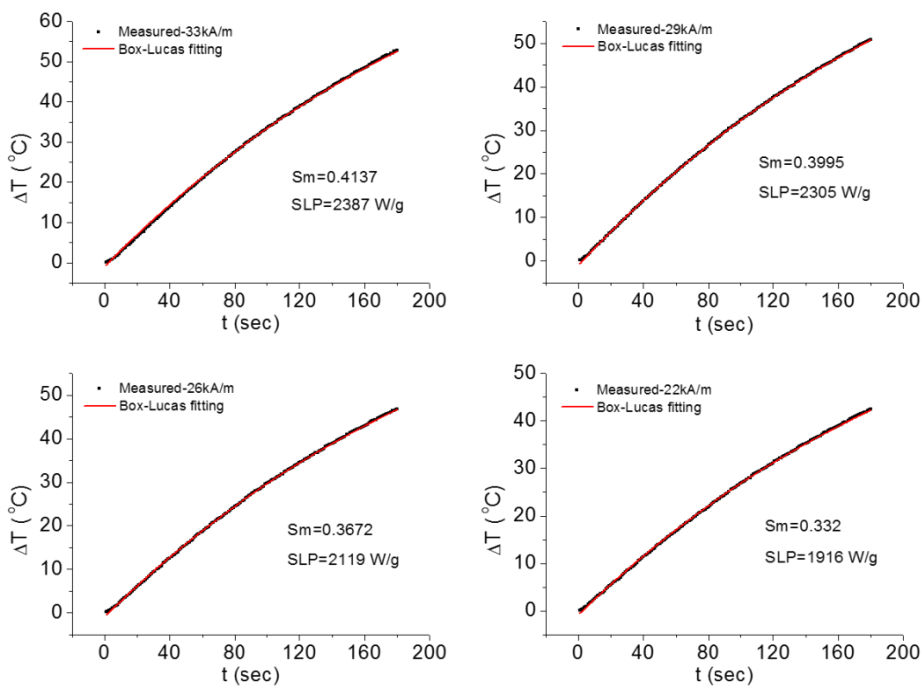

Figure S9. Examples of heating curve (black) and Box-Lucas fitting curve (red) of $Zn_{0.3}Fe_{2.7}O_4/SiO_2$ NP solution exposed in the AC field of 380 kHz under different field amplitudes.

## 9. SLP and ILP of NPs in recent studies

| Sample | SLP (W/g$_{metal}$) | ILP (nHm$^2$/kg) | AC field | Reference |
|---|---|---|---|---|
| Multicore γ-Fe$_2$O$_3$ | 2000 | 4.6 | 700 kHz, 25 kA/m | Ref. S3 |
| Zn$_{0.4}$Co$_{0.6}$Fe$_2$O$_4$/ Zn$_{0.4}$Mn$_{0.6}$Fe$_2$O$_4$ | 3886 | 5.6 | 500 kHz, 37.3 kA/m | Ref. S4 |
| Fe$_3$O$_4$ | 2452 | 5.6 | 520 kHz, 29 kA/m | Ref. S5 |
| Fe$_3$O$_4$ | 1000 | 6.3 | 100 kHz, 40 kA/m | Ref. S6 |
| Fe$_3$O$_4$ | 332 | 8.3 | 400 kHz, 10 kA/m | Ref. S7 |
| magnetosomes | 960 | 23.4 | 410 kHz, 10 kA/m | Ref. S8 |
| Co$_{0.03}$Mn$_{0.27}$Fe$_{2.7}$O$_4$ | 3417 | 8.3 | 380 kHz, 33 kA/m | This work |
| Zn$_{0.3}$Fe$_{2.7}$O$_4$ | 1010 | 15.7 | 380 kHz, 13 kA/m | This work |
| Zn$_{0.3}$Fe$_{2.7}$O$_4$ | 500 | 26.7 | 380 kHz, 7 kA/m | This work |
| Zn$_{0.3}$Fe$_{2.7}$O$_4$ | 282 | 59.9 | 380 kHz, 3 kA/m | This work |

## 10. Measurements of the SLP at different percentage of glycerol

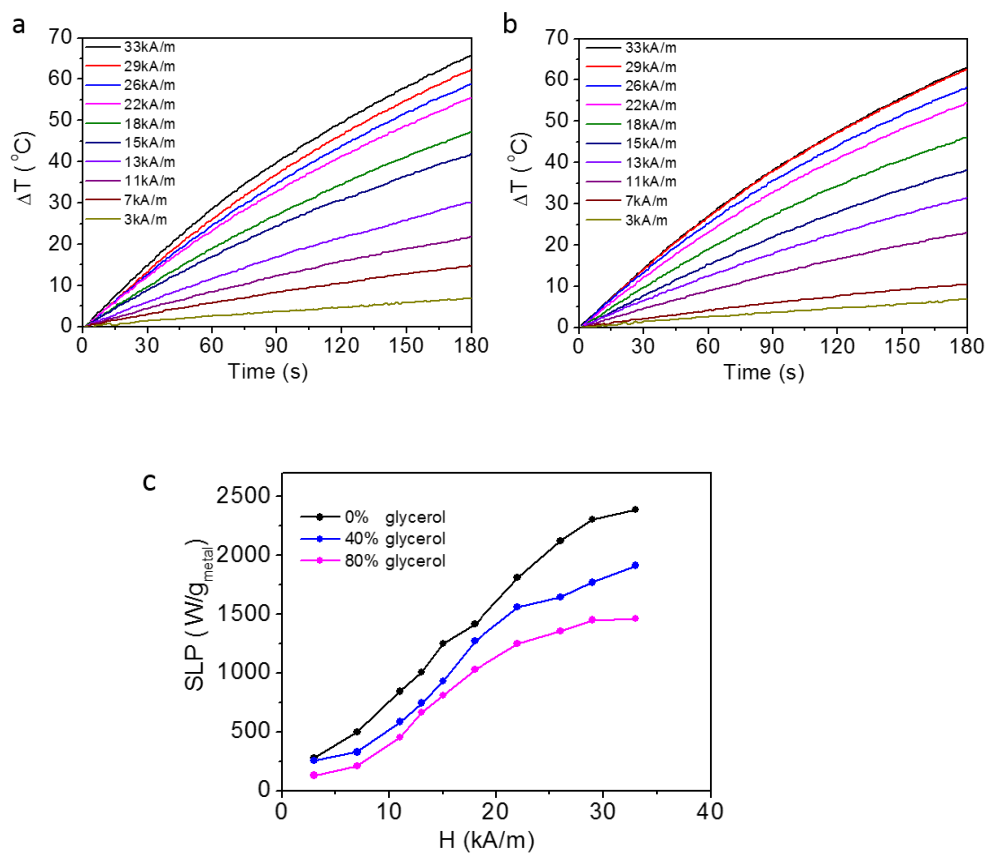

Figure S10. AC field heating of NPs measured by temperature *vs* time curves for $Zn_{0.3}Fe_{2.7}O_4/SiO_2$ NP water/glycerol mixture dispersions (1 $mg_{NPs}$/ml) (a) containing 40% glycerol, (b) containing 80% glycerol, under an AC field of 380 kHz with different field amplitudes; (c) Field dependence of SLP for $Zn_{0.3}Fe_{2.7}O_4/SiO_2$ aqueous and water/glycerol mixture dispersions.

## 11. AC field induced cell apoptosis (cell death) with different exposure time

The cell apoptosis for MG-63 cells incubated with ZFO NPs and exposed to an ac magnetic field with different exposure time, as shown in Fig.S11. In Fig S11 (c) to (f), the third quadrant demonstrates the early apoptosis of cell, and the fourth quadrant represents late apoptosis. The total cell apoptosis is the sum of early apoptosis and late apoptosis. The efficiency of cell apoptosis increases with increasing exposure time in AC magnetic field. In particular, the percentage of late apoptotic cells increases more prominently with increasing exposure time. Magnetic hyperthermia at ac field of 380 kHz and 13 kA/m for 30 min leads to about 89% cell apoptosis.

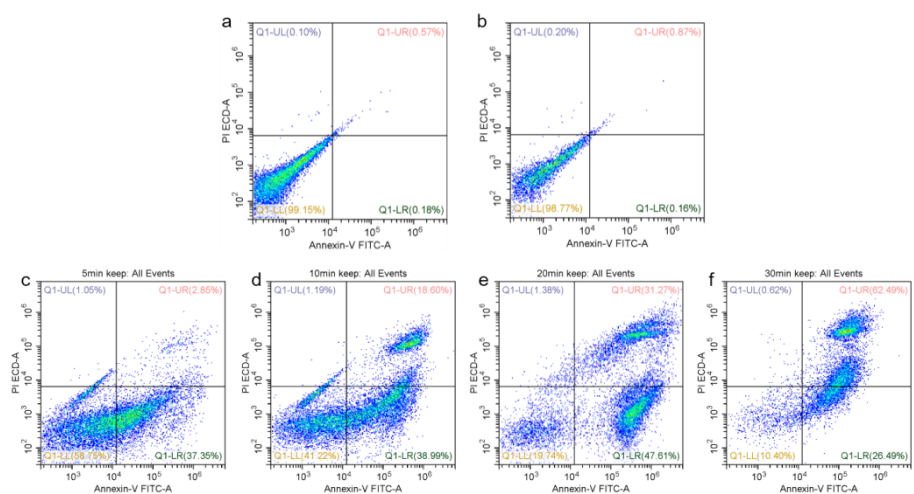

Figure S11. Apoptosis assay fluorescence from Annexin V and PI uptake by the MG-63 cells were monitored by flow cytometry. (a)MG-63 cells without NPs used as a control group.(b) MG-63 cells with NPs but not exposed to ac field used as another control group; MG-63 cells incubated with 300 μg/ml NPs heated under the ac field of 380 kHz, 13 kA m$^{-1}$ for (c) 5 min (d) 10 min (e) 20 min (f) 30 min.